\documentclass[aps,pre,preprint,showpacs,amsmath,amsfonts,preprintnumbers]{revtex4}
\usepackage{epsfig}
\usepackage{yfonts}
\usepackage{amssymb}
\usepackage{graphicx}
\swabfamily
\newcommand{\Op}[1]{{{\mathrm{\hat{#1}}}}}

\begin{document}
\title{The globally stable solution of a stochastic Nonlinear Schr\"odinger Equation.}
\author{M. Khasin and R. Kosloff}
\affiliation{Fritz Haber Research Center for Molecular Dynamics, 
Hebrew University of Jerusalem, Jerusalem 91904, Israel}
\date{\today }

\begin{abstract}
Weak measurement of a subset of noncommuting observables of a quantum system can be modeled by the open-system evolution, governed by the master equation in the Lindblad form \cite{diosi06}. The open-system density operator can be represented as statistical mixture over non unitarily evolving pure states, driven by the stochastic Nonlinear Schr\"odinger equation (sNLSE) \cite{gisin84,diosi88,gisin92}.
The globally stable solution of the sNLSE is obtained in the case where the measured subset of observables comprises the spectrum-generating algebra of the system. This solution is a generalized coherent state (GCS)\cite{Perelomov,Gilmore}, associated with the algebra. The result is based on proving that  GCS minimize the trace-norm of the covariance matrix, associated with the spectrum-generating algebra.

\end{abstract}
\pacs{03.67.Mn,03.67.-a, 03.65.Ud, 03.65 Yz}
\maketitle
\section{Introduction}
A weak measurement of an subset of operators $\left\{\Op X_i\right\}_{i=1}^K$, performed on a quantum system, driven by the Hamiltonian  $\Op H$ is described by a Lindblad master equation \cite{lindblad76,breuer} of the following form:
\begin{eqnarray}
\frac{\partial}{\partial t}\Op \rho={\cal L} \Op \rho=-i \left[\Op H, \Op\rho \right]- \sum_{j=1}^K  \gamma_j\left[\Op X_j,\left[\Op X_j,\Op\rho \right]\right], \label{liouville}
\end{eqnarray} 
where $\gamma_j$, measures the strength of the measurement of the observable $\Op X_j$. The solution of the master equation (\ref{liouville}) can be represented as a statistical mixture of pure-states, evolving, according to 
the stochastic nonlinear Schr\"odinger equation (sNLSE)\cite{gisin84,diosi88,gisin92}: 
\begin{eqnarray}
d\left|{\psi}\right\rangle &=&\left\{-i \Op H dt - \sum_{i=1}^K  \gamma_j \left( \Op X_i-\left\langle \Op X_i\right\rangle_{\psi}  \right)^2 dt+ \sum_{i=1}^K\left(  \Op X_i-\left\langle \Op X_i\right\rangle_{\psi}\right) d\xi_i\right\}\left|{\psi}\right\rangle, \label{snlse}
\end{eqnarray}
where the Wiener fluctuation terms $d\xi_i$ satisfy 
\begin{eqnarray}
<d\xi_i>=0, \ \ \ d\xi_id\xi_j=2\gamma_j dt.
\end{eqnarray}

Consider a quantum system, driven by the Lie-algebraic Hamiltonians:
\begin{eqnarray}
\Op H=\sum_j a_j \Op X_j, \label{laham}
\end{eqnarray}
where the set $\{ \Op X_j \}$ of observables is closed under the commutation relations:
\begin{eqnarray}
\left[\Op X_i,\Op X_j\right]=i \sum_{k=1}^K  f_{ijk}\Op X_k ,\label{commutation}
\end{eqnarray}
i.e., it forms the spectrum-generating \cite{bohm} Lie algebra \cite{gilmorebook}  of the system, labeled by the letter $\mathfrak g$ in what follows. The coefficients $a_j$ in the Hamiltonian (\ref{laham}) can be functions of the Casimir operators \cite{gilmorebook} of the algebra.  Such Lie-algebraic Hamiltonians (\ref{laham}) are encountered in various fields of the many-body physics, such as  molecular \cite{Iachello, iachello06}, nuclear \cite{bohm, iachello06} and condensed matter physics \cite{bohm}. 
Lie algebras considered in the present work are compact semisimple algebras \cite{gilmorebook} and the basis $\{ \Op X_i \}$ is assumed to be  orthonormal   with respect to the Killing form \cite{gilmorebook}. 

We shall assume that a weak measurement is performed on the elements of the spectrum-generating algebra $\mathfrak g$. The strengts of the all the measurements are assumed to be equal, $\gamma_j=\gamma$.
Our goal is finding globally stable solutions of the resulting sNLSE (\ref{snlse}), which can be interpreted as a single realization of an infinite  series of weak measurements.

\section{Generalized coherent states and the total uncertainty}
Let us assume that the subalgebra $\mathfrak g$ is represented irreducibly on the system's Hilbert space $\cal H$. Then an arbitrary state $\psi\in \cal H$ can be represented as a superposition of the \textit{generalized coherent states (GCS)} \cite{Perelomov,Gilmore} $\left|\Omega,\psi_0\right\rangle$ with respect to the corresponding dynamical group $G$ and an arbitrary state $\psi_0$:
\begin{eqnarray}
\left|\psi\right\rangle=\int d \mu(\Omega)\left|\Omega,\psi_0\right\rangle\left\langle \Omega, \psi_0|\psi\right\rangle,  \label{expansion1}
\end{eqnarray}
where $\mu(\Omega)$ is the group invariant measure on the coset space $G/H$ \cite{gilmorebook} , $\Omega\in G/H$,  $H\subset G$ is the maximal stability subgroup of the reference state $\psi_0$:
\begin{eqnarray}
h\left|\psi_0\right\rangle=e^{i\phi(h)}\left|\psi_0\right\rangle, \ \ h\in H
\end{eqnarray}
and the GCS $\left|\Omega,\psi_0\right\rangle$ are defined as follows:
\begin{eqnarray}
\Op U(g)\left|\psi_0\right\rangle=\Op U(\Omega h)\left|\psi_0\right\rangle=e^{i\phi(h)}\Op U(\Omega ) \left|\psi_0\right\rangle\equiv e^{i\phi(h)} \left|\Omega, \psi_0\right\rangle, \ \ g\in G, \ h\in H, \ \Omega\in G/H, \label{gcs}
\end{eqnarray}
where $\Op U(g)$ is a unitary transformation generated by a group element $g\in G$.

The  group-invariant \textit{total uncertainty} of a state with respect to a compact semisimple algebra $\mathfrak g$ is defined as \cite{delbourgo, Perelomov}:
\begin{eqnarray}
\Delta[\psi]\equiv\sum_{j=1}^K \left\langle \Delta \Op X_j^2\right\rangle_{\psi}=\sum_{j=1}^K  \left\langle  \Op X_j^2\right\rangle_{\psi}- \sum_{j=1}^K \left\langle  \Op X_j\right\rangle_{\psi}^2\label{giu}.
\end{eqnarray}
The first term in the rhs of Eq.(\ref{giu}) is the eigenvalue  of the   the Casimir operator  of $\mathfrak g$ in the Hilbert space representation:
\begin{eqnarray}
\Op C=\sum_{j=1}^K \Op X^2_j \label{Casimir} 
\end{eqnarray}
and the second term is termed the generalized purity \cite{Viola03} of the state with respect to $\mathfrak g$:
\begin{eqnarray}
P_{\mathfrak g}[\psi]\equiv \sum_{j=1}^K \left\langle  \Op X_j\right\rangle_{\psi}^2 \label{gpurity}.
\end{eqnarray}
Let us define $\Delta_{min}$ as a minimal total uncertainty of a quantum state  and  $c_{\cal H}$ as the eigenvalue  of the   the Casimir operator  of $\mathfrak g$ in the system Hilbert space. Then
\begin{eqnarray}
\Delta_{min} \le \Delta[\psi]\le c_{\cal H}, \label{giu1}
\end{eqnarray}

The total uncertainty (\ref{giu}) is invariant under an arbitrary unitary transformation generated by $\mathfrak g$. Therefore,  all the GCS with respect to the subalgebra $\mathfrak g$ and a reference state $\psi_0$ have a fixed value of the total invariance. It has been proved in Ref.\cite{delbourgo} that 
the minimal total uncertainty $\Delta_{min}$ is obtained if and only if $\psi_0$ is a highest (or lowest) weight state of the representation (the Hilbert space). The value of  $\Delta_{min}$ is given by \cite{delbourgo, klyachko} 
\begin{eqnarray}
\Delta_{min}\equiv (\Lambda,\mu) \le \Delta[\psi]\le (\Lambda,\Lambda+\mu)=c_{\cal H}, \label{giu2}
\end{eqnarray}
where $\Lambda \in {\mathbb R}^r$ is the the highest weight of the representation, $\mu \in {\mathbb R}^r$ is the  sum of  the positive roots of $\mathfrak g$,  $r$ is the rank of $\mathfrak g$ \cite{gilmorebook} and $(,)$ is the Euclidean scalar product in ${\mathbb R}^r$.
The corresponding CGS were termed the generalized unentangled states with respect to the subalgebra $\mathfrak g$ \cite{Viola03, klyachko}. 
The maximal value  of the uncertainty is obtained in states termed maximally or completely entangled  \cite{Viola03, klyachko} with respect to $\mathfrak g$. The maximum value equals  $c_{\cal H}$ in the states having $\left\langle \psi\left|\Op X_j\right|\psi \right\rangle^2=0$ for all $i$. Such states exist in a generic irreducible representation of an arbitrary compact simple algebra of observables \cite{klyachko}. Generic superpositions of the GCS have larger uncertainty and are termed generalized entangled states with respect to $\mathfrak g$ \cite{Viola03, klyachko}.
In what follows, it is assumed that the reference state $\psi_0$  for the GCS minimize the total invariance (\ref{giu}). 

\section{The main result: global stability of the generalized coherent states}
The time evolution of the total uncertainty (\ref{giu}) of a pure state evolving according to the sNLSE (\ref{snlse}) can be calculated as follows:
\begin{eqnarray}
d  {\Delta}[\psi(t)]&=&d \sum_i  \left( \left\langle \Op X_i^2\right\rangle_{\psi}-\left\langle \Op X_i\right\rangle_{\psi}^2\right)=-d \sum_i \left\langle \Op X_i\right\rangle_{\psi}^2\nonumber \\
&=&- \sum_i  \left(2d\left\langle \Op X_i\right\rangle_{\psi}\left\langle \Op X_i\right\rangle_{\psi}+d\left\langle \Op X_i\right\rangle_{\psi}d\left\langle \Op X_i\right\rangle_{\psi}\right), \label{differential}
\end{eqnarray}
where we have used prescription of the Ito calculus: $d(xy)=dx y+x dy+dx dy$ and the fact that $d \sum_i  \left\langle \Op X_i^2\right\rangle_{\psi}=0$
by the invariance of the Casimir operator (\ref{Casimir}) under dynamics in an irreducible representation.
To calculate $d\Op X_i$ we derive the Heisenberg equations of motion, corresponding to  the sNLSE (\ref{snlse}).

 Eq.(\ref{snlse}) is equivalent to the following equation for the corresponding projector $\Op P_{\psi}=\left|\psi\right\rangle\left\langle \psi\right|$ 
\begin{eqnarray}
d \Op P_{\psi}=\left(-i \left[\Op H, \Op P_{\psi} \right]-\gamma \sum_{j=1}^K  \left[\Op X_j,\left[\Op X_j,\Op P_{\psi} \right]\right]\right)dt+\sum_i  \left\{\left( \Op X_i-\left\langle \Op X_i\right\rangle_{\psi}   \right)d \xi_i,\Op P_{\psi}\right\}. \label{stochliouville}
\end{eqnarray} 
Eq.(\ref{stochliouville}) implies the following stochastic Heisenberg equation for an arbitrary operator $\Op X_i$:
\begin{eqnarray}
d \Op X_i&=&\left(i \left[\Op H, \Op X_i \right]-\gamma \sum_{j=1}^K  \left[\Op X_j,\left[\Op X_j,\Op X_i \right]\right]\right)dt+\sum_j \left\{\left( \Op X_j-\left\langle \Op X_j\right\rangle_{\psi}   \right)d \xi_j,\Op X_i \right\}\nonumber \\
&=&\left(i \left[\Op  H, \Op X_i \right]-\gamma c_{\texttt{adj}} \Op X_i \right)dt+\sum_j \left\{\left( \Op X_j-\left\langle \Op X_j\right\rangle_{\psi}   \right)d \xi_j,\Op X_i \right\},
 \label{stochheisenberg}
\end{eqnarray}  
where $c_{\texttt{adj}}$ is the quadratic Casimir in the adjoint representation (see Eq.(\ref{Casimir})).
Multiplying Eq.(\ref{stochheisenberg}) by $\left\langle \Op X_i\right\rangle_{\psi}$, summing up over all the observables and computing the expectation value we obtain
\begin{eqnarray}
\sum_{i=1}^K \left\langle \Op X_i\right\rangle_{\psi} d \left\langle \Op X_i\right\rangle
&=&\left(i \sum_{i=1}^K \left\langle \Op X_i\right\rangle_{\psi} \left\langle \left[\Op  H, \Op X_i \right]\right\rangle_{\psi}-\gamma c_{\texttt{adj}} \sum_{i=1}^K \left\langle \Op X_i\right\rangle_{\psi}^2 \right)dt\nonumber \\
&+&\sum_{j,i}\xi_j \left\langle \Op X_i\right\rangle_{\psi} \left\langle \left\{\left( \Op X_j-\left\langle \Op X_j\right\rangle_{\psi} \right)d ,\Op X_i \right\}\right\rangle_{\psi}\nonumber \\
&=& -\gamma c_{\texttt{adj}} \sum_{i=1}^K \left\langle \Op X_i\right\rangle_{\psi}^2 dt\nonumber \\
&+&\sum_{i,j=1}^K  \left\langle \Op X_i\right\rangle_{\psi} \left( \left\langle \left\{\Op X_j,\Op X_i \right\}\right\rangle_{\psi}-2\left\langle \Op X_j\right\rangle_{\psi}\left\langle \Op X_i\right\rangle_{\psi}\right)d\xi_j,
 \label{stochheisenberg1}
 \end{eqnarray}
 where the contribution of the  Hamiltonian term has vanished due to the antisymmetry of the structure constants of $\mathfrak g$:
 \begin{eqnarray}
i \sum_{i,j=1}^K a_j\left\langle \Op X_i\right\rangle_{\psi} \left\langle \left[\Op  X_j, \Op X_i \right]\right\rangle_{\psi}&=&i \sum_{i,j,k=1}^K a_j\left\langle \Op X_i\right\rangle_{\psi} \left\langle ic_{jik}\Op  X_k \right\rangle_{\psi}\nonumber \\
&=&-\sum_{j=1}^K a_j\sum_{i,k=1}^K c_{jik}\left\langle \Op X_i\right\rangle_{\psi} \left\langle \Op  X_k \right\rangle_{\psi}=0.
\end{eqnarray}
 
 From  Eq.(\ref{stochheisenberg}) we get 
\begin{eqnarray}
d\left\langle \Op X_i\right\rangle_{\psi}d\left\langle \Op X_i\right\rangle_{\psi}&=&\sum_{k,l}d\xi_k d\xi_l\left\langle \left\{\left( \Op X_k-\left\langle \Op X_k\right\rangle_{\psi}   \right) ,\Op X_i \right\}\right\rangle_{\psi} \left\langle \left\{\left( \Op X_l-\left\langle \Op X_l\right\rangle_{\psi}   \right),\Op X_i \right\}\right\rangle_{\psi} \nonumber \\
&=&2\gamma dt \sum_{k}\left\langle \left\{\left( \Op X_k-\left\langle \Op X_k\right\rangle_{\psi}   \right) ,\Op X_i \right\}\right\rangle_{\psi}^2\nonumber \\
&=&2\gamma dt \sum_{k} \left( \left\langle \left\{\Op X_k,\Op X_i \right\}\right\rangle_{\psi}-2\left\langle \Op X_k\right\rangle_{\psi}\left\langle \Op X_i\right\rangle_{\psi}\right)^2.
 \label{stochheisenberg3}
\end{eqnarray} 
Inserting Eqs.(\ref{stochheisenberg3}) and (\ref{stochheisenberg1}) into Eq.(\ref{differential}) we obtain 
\begin{eqnarray}
d\left\langle \Op {\Delta}\right\rangle_{\psi}
&=&- \sum_i  \left(2d\left\langle \Op X_i\right\rangle_{\psi}\left\langle \Op X_i\right\rangle_{\psi}+d\left\langle \Op X_i\right\rangle_{\psi}d\left\langle \Op X_i\right\rangle_{\psi}\right)\nonumber \\
&=&2\gamma  \left( c_{\texttt{adj}} \sum_{i=1}^K \left\langle \Op X_i\right\rangle_{\psi}^2 -  \sum_{k,i} \left( \left\langle \left\{\Op X_k,\Op X_i \right\}\right\rangle_{\psi}-2\left\langle \Op X_k\right\rangle_{\psi}\left\langle \Op X_i\right\rangle_{\psi}\right)^2\right)dt\nonumber \\
&-&2\sum_{j,i} \left\langle \Op X_i\right\rangle_{\psi} \left( \left\langle \left\{\Op X_j,\Op X_i \right\}\right\rangle_{\psi}-2\left\langle \Op X_j\right\rangle_{\psi}\left\langle \Op X_i\right\rangle_{\psi}\right)d\xi_j . \label{differential1}
\end{eqnarray}
The remaining terms in the Eq.(\ref{differential1})  describe the effect of the bath (weak measurement) on the total uncertainty of a pure state evolving according to the sNLSE. It can be shown by direct calculation that these terms vanish in a GCS. But a simpler way to show this is to note that the infinitesimal evolution of the state, corresponding to the sNLSE (\ref{snlse}) dropping the Hamiltonian term,  is given by:
\begin{eqnarray}
|\psi>+|d\psi>&=&\exp\left\{-2\gamma \Op \Delta+ \sum_i\left(  \Op X_i-\left\langle \Op X_i\right\rangle_{\psi}\right) d\xi_i\right\}|\psi>\nonumber \\
&=&\exp\left\{\sum_i\left(  \Op X_i-\left\langle \Op X_i\right\rangle_{\psi}\right) d\xi_i\right\} \exp\left\{-2\gamma\Op \Delta\right\}|\psi>\nonumber \\
&=&\exp\left\{\phi(t)\right\} \exp\left\{\sum_i\left(  \Op X_i-\left\langle \Op X_i\right\rangle_{\psi}\right) d\xi_i\right\} |\psi>, \label{infi}
\end{eqnarray} 
where we have used the notation $\Op \Delta \equiv \sum_i  \left( \Op X_i-\left\langle \Op X_i\right\rangle_{\psi}   \right)^2$  and the fact \cite{delbourgo} that a GCS is an eigenstate of $\Op \Delta$. From 
Eq.(\ref{infi}) we see that the infinitesimal transformation of the state is driven by the operator \textit{linear} in the generators of the algebra. Therefore, a GCS transforms into a GCS under the infinitesimal evolution \footnote{This fact does not follow directly from the definition of the GCS, since the evolution in Eq.(\ref{infi}) is not unitary, but is nonetheless correct, see \cite{Gilmore}.} and the total uncertainty of the evolving state remains constant (and minimal).

The first term in Eq.(\ref{differential1}), considered as a functional on the Hilbert space, has global maximum in the GCS (see below). Therefore, on average, the rate of localization is minimal in a GCS. In a GCS the second (stochastic) term vanishes.  Since the rate of localization is zero in a GCS as proved above, it follows that the average rate of localization obtains minimum at zero. Therefore, an arbitrary state localizes on average. 

Next we prove that the first term in Eq.(\ref{differential1}), considered as a functional on the Hilbert space, has global maximum in the GCS.
The first sum in this term  is just the generalized purity of the state (\ref{gpurity}), which has a global maximum in a GCS \cite{Viola03, Klyachko08}, while the second sum is the trace-norm of the covariance matrix, which  obtains global minimum in a GCS. 

\textbf{Theorem.} \textit{The trace-norm of the covariance matrix $M_{ij}= \left\langle \left\{\Op X_k,\Op X_i \right\}\right\rangle_{\psi}-2\left\langle \Op X_k\right\rangle_{\psi}\left\langle \Op X_i\right\rangle_{\psi}$ is minimal in a maximal (minimal) weight state of the irrep, i.e., in a GCS.}

\textit{Proof:} The trace-norm is invariant under unitary transformations, generated by the algebra $\mathfrak g$. Therefore, any orthonormal basis $\Op X_i$ can be used for calculation of the trace-norm. Consider particular choice of the basis $\Op X_i$ such that the projection of the  pure state $\Op \rho=\left|\psi\right\rangle\left\langle \Op \psi\right|$ on $\mathfrak g$ is contained in the Cartan subalgebra $\mathfrak h \subset \mathfrak g$.
Let us use index $i,j$ for the elements of $\mathfrak h $ and $\alpha, \beta$ for the elements of the root subspace. Then,
\begin{eqnarray}
\texttt{Tr} \{ M^2 \}&=& \sum_{i,j} M_{i,j}^2+\sum_{i,\alpha} M_{i,\alpha}^2+\sum_{|\alpha|\neq|\beta|} M_{\alpha,\beta}^2+\sum_{|\alpha|=|\beta|} M_{\alpha,\beta}^2. \label{sums}
\end{eqnarray}
Let us focus on the last term in Eq.(\ref{sums}). Since the projection of the state on  $\mathfrak g$ is contained in the Cartan subalgebra, it vanishes on the root subspace, i.e., $\left\langle \Op X_{\alpha}\right\rangle=0$, for every $\alpha$. Then
\begin{eqnarray}
\sum_{|\alpha|=|\beta|} M_{\alpha,\beta}^2=\sum_{|\alpha|=|\beta|}\left\langle \left\{\Op X_\alpha,\Op X_\beta \right\}\right\rangle_{\psi}^2
\end{eqnarray}
Using notation $E_{\pm \alpha}$ for the raising and the lowering operators of the algebra, corresponding to the positive root $\alpha$ we obtain
\begin{eqnarray}
&&\sum_{|\alpha|=|\beta|} M_{\alpha,\beta}^2 \nonumber  \\
&=&\sum_{|\alpha|=|\beta|}\left\langle \left\{\Op X_\alpha,\Op X_\beta \right\}\right\rangle_{\psi}^2=-\frac{1}{2}\sum_{\alpha>0}\left\langle \left(\Op E_{\alpha}+\Op E_{-\alpha}\right) \left(\Op E_{\alpha}-\Op E_{-\alpha}\right)+ \left(\Op E_{\alpha}-\Op E_{-\alpha}\right) \left(\Op E_{\alpha}+\Op E_{-\alpha}\right)   \right\rangle^2 \nonumber \\
&+&\sum_{\alpha>0}\left\langle \left(\Op E_{\alpha}+\Op E_{-\alpha}\right)^2   \right\rangle^2+\sum_{\alpha>0}\left\langle \left(\Op E_{\alpha}-\Op E_{-\alpha}\right)^2   \right\rangle^2=-2\sum_{\alpha>0}\left\langle \Op E_{\alpha}^2-\Op E_{-\alpha}^2   \right\rangle^2 \nonumber \\
&+&\sum_{\alpha>0}\left\langle \Op E_{\alpha}^2+\Op E_{-\alpha}^2 + \Op E_{\alpha} \Op E_{-\alpha}+\Op E_{-\alpha} \Op E_{\alpha}\right\rangle^2+\sum_{\alpha>0}\left\langle \Op E_{\alpha}^2+\Op E_{-\alpha}^2 - \Op E_{\alpha} \Op E_{-\alpha}-\Op E_{-\alpha} \Op E_{\alpha}\right\rangle^2 \nonumber \\
&=&-2\sum_{\alpha>0}\left\langle \Op E_{\alpha}^2-\Op E_{-\alpha}^2   \right\rangle^2+2\sum_{\alpha>0}\left\langle \Op E_{\alpha}^2+\Op E_{-\alpha}^2   \right\rangle^2+2\sum_{\alpha>0}\left\langle \Op E_{\alpha} \Op E_{-\alpha}+\Op E_{-\alpha} \Op E_{\alpha}   \right\rangle^2 \nonumber \\
&=& 8\sum_{\alpha>0}\left\langle \Op E_{\alpha}^2\right\rangle\left\langle \Op E_{-\alpha}^2   \right\rangle+2\sum_{\alpha>0}\left\langle \Op E_{\alpha} \Op E_{-\alpha}+\Op E_{-\alpha} \Op E_{\alpha}   \right\rangle^2.  \label{rootexp}
\end{eqnarray}

The density operator $\Op \rho=\left|\psi\right\rangle\left\langle \psi\right|$ can be expressed in the basis of the eigenstates $\left|\mu\right\rangle$ of the Cartan operators $\Op X_i\in \mathfrak h $,  $\Op X_i\left|\mu \right\rangle=\mu_i\left|\mu \right\rangle$
\begin{eqnarray}
\Op \rho=\sum_{\mu,\mu'}c_{\mu}c_{\mu'}^*\left|\mu\right\rangle\left\langle \mu'\right|.
\end{eqnarray}
Then the  the last term in Eq.(\ref{rootexp}) obtains
\begin{eqnarray}
2\sum_{\alpha>0}\left\langle \Op E_{\alpha} \Op E_{-\alpha}+\Op E_{-\alpha} \Op E_{\alpha}   \right\rangle^2&=&2\sum_{\alpha>0}\left(\sum_{\nu,\mu'}c_{\nu}c_{\mu'}^*\left\langle \mu'\right| \Op E_{\alpha} \Op E_{-\alpha}+\Op E_{-\alpha} \Op E_{\alpha}  \left|\nu\right\rangle \right)^2 \nonumber \\
&=&2\sum_{\alpha>0}\left(\sum_{\nu}|c_{\nu}|^2\left\langle \nu \right| \Op E_{\alpha} \Op E_{-\alpha}+\Op E_{-\alpha} \Op E_{\alpha}  \left|\nu\right\rangle \right)^2. \label{mainterm}
\end{eqnarray}
States $\left|\nu+k \alpha \right\rangle$ form an irreducible representation of the ${\mathfrak {su}}(2)$, spanned by 
\begin{eqnarray}
E^{\pm}&\equiv& E_{\pm \alpha}/|\alpha| \nonumber \\
E_{3}&\equiv& \alpha \cdot \Op H/|\alpha|^2, \ \ \Op H_i\equiv\Op X_i\in \mathfrak h  \label{su2def}
\end{eqnarray}
obeying ${\mathfrak {su}}(2)$ commutation relations \cite{georgi}
\begin{eqnarray}
[E_{3}, E^{\pm}]=\pm E^{\pm}; \ \ [E^{+}, E^{-}]=\pm E^{\pm}.
\end{eqnarray}
Therefore, the state $\left|\nu \right\rangle$ can be labeled as $\left|m_{\alpha},j_{\alpha} \right\rangle$, where  $j_{\alpha}$ is the maximal weight of the corresponding irrep of the ${\mathfrak {su}}(2)$ and $m_{\alpha}$ is the weight, corresponding to the state $\left|\nu \right\rangle$ in the irrep. Then
\begin{eqnarray}
\left\langle \Op E_{-\alpha}\Op E_{\alpha}+ \Op E_{\alpha}\Op E_{-\alpha}\right\rangle_{\psi}^2&=&|\alpha|^4\left\langle 2 \Op E^{-}\Op E^{+}+ \Op E_{3}\right\rangle_{\psi}^2=|\alpha|^4\left\langle m_{\alpha},j_{\alpha}\left|2 \Op E^{-}\Op E^{+}+ \Op E_{3}\right|m_{\alpha},j_{\alpha}\right\rangle^2 \nonumber \\
&=&|\alpha|^4\left( j_{\alpha}+ j_{\alpha}^2-m_{\alpha}^2 \right)^2. \label{single}
\end{eqnarray}
The term (\ref{single}) obtains minimum in the maximal (minimal) weight state of the $j_{\alpha}$ irrep, corresponding to $m_{\alpha}=j_{\alpha} (m_{\alpha}=-j_{\alpha})$. Therefore,  $\left|m_{\alpha},j_{\alpha}  \right\rangle$ is annihilated by the $E^+(E^-)$, and, by Eqs.(\ref{su2def}), 
the state $\left|\nu \right\rangle$ is annihilated by $E_{\alpha}(E_{-\alpha})$. The minimum of the sum (\ref{mainterm}) is obtained in the state, annihilated by $E_{\alpha}(E_{-\alpha})$ for all positive roots $\alpha$, i.e., in the maximal (minimal) weight state $\Op \rho=\left|\Lambda\right\rangle\left\langle \Lambda\right|$. The    first term in Eq.(\ref{rootexp}) is nonnegative and vanishes at $\left|\psi\right\rangle=\left|\Lambda\right\rangle$, therefore it obtains minimum at $\left|\Lambda\right\rangle$. Therefore, the term (\ref{rootexp}) in the sum (\ref{sums}) obtains minimum at $\left|\Lambda\right\rangle$. Since $\left|\Lambda\right\rangle$ is an eigenstate of every Cartan operator $\Op X_i$, the first term in Eq.(\ref{sums}) vanishes at  $\left|\Lambda\right\rangle$. For the same reason and the fact that projection of $\Op \rho$ on the root subspace vanishes the second term in Eq.(\ref{sums}) also vanishes at  $\left|\Lambda\right\rangle$. The third term in Eq.(\ref{sums}) vanishes at $\left|\Lambda\right\rangle$ since $\left\langle \Lambda\left|\Op E_{\alpha}\Op E_{\beta}\right|\Lambda\right\rangle=0$, $\forall |\alpha|\neq |\beta|$. Since all these terms are nonnegative, they obtain minimum at the maximal (minimal) weight state $\Op \rho=\left|\Lambda\right\rangle\left\langle \Lambda\right|$. Therefore, the whole expression (\ref{sums}) for the trace norm of the covariance matrix obtains minimum at  $\Op \rho=\left|\Lambda\right\rangle\left\langle \Lambda\right|$. $\Box$
 
 \section{Conclusions}
 It is proved that globally stable solutions of the stochastic Nonlinear Schr\"odinger Equation (\ref{snlse}), modeling the process of weak measurement of the elements of the spectrum-generating algebra of the system,  are generalized coherent states, associated with the algebra. The Hamiltonian of the system is linear in the algebra elements, i.e., possess dynamical symmetry. It is conjectured, that adding nonlinearity to the Hamiltonian results in the asymptotically stable localized solutions of the corresponding sNLSE (see Ref.\cite{Khasin081} for some numerical evidence). The proof of stability is based on proving that the trace-norm of the covariance matrix, associated with the algebra, obtains minimum in a generalized coherent state.

\begin{acknowledgments}
Work supported by DIP and the Israel Science Foundation (ISF).
The Fritz Haber Center is supported
by the Minerva Gesellschaft f\"{u}r die Forschung GmbH M\"{u}nchen, Germany.
\end{acknowledgments}

 \end{document}